\documentclass[
    ,final            
  ]
  {aipproc}
\usepackage[lflt]{floatflt}
\usepackage{float}
\usepackage{wrapfig}
\layoutstyle{6x9}

\begin{document}
\title{The TANAMI Program}

\author{Cornelia M\"uller}{
  address={Dr. Remeis-Sternwarte \& ECAP, Sternwartstrasse 7, 96049 Bamberg, Germany}
}

\author{Matthias Kadler}{
  address={Dr. Remeis-Sternwarte \& ECAP, Sternwartstrasse 7, 96049 Bamberg, Germany}
,altaddress={CRESST/NASA Goddard Space Flight Center, Greenbelt, MD 20771, USA \&
	\\USRA, 10211 Wincopin Circle, Suite 500 Columbia, MD 
             21044, USA}
}

\author{Roopesh Ojha}{
  address={USNO, 3450 Massachusetts Ave., NW, Washington DC 20392, USA},
,altaddress={NVI, Inc., 7257D Hanover Parkway, Greenbelt, MD 20770, USA}
}

\author{M. B\"ock}{
  address={Dr. Remeis-Sternwarte \& ECAP, Sternwartstrasse 7, 96049 Bamberg, Germany}
}

\author{\mbox{the TANAMI Team}}{
  address={see full author list on page 3}
}

\begin{abstract}
TANAMI (Tracking Active Galactic Nuclei with Austral Milliarcsecond Interferometry) is a monitoring program to study the parsec-scale structures and dynamics of relativistic jets in active galactic nuclei (AGN) of the Southern Hemisphere with the Long Baseline Array and associated telescopes. Extragalactic jets south of $-30^\circ$ declination are observed at $8.4\,$GHz and $22\,$GHz every two months at milliarcsecond resolution. The initial TANAMI sample is a hybrid radio and $\gamma$-ray selected sample since the combination of VLBI and $\gamma$-ray observations is crucial to understand the broadband emission characteristics of AGN.
\end{abstract}

\maketitle
TANAMI (Tracking Active Galactic Nuclei with Austral Milliarcsecond Interferometry) observations are made with the five antennas of the Australian Long Baseline array, telescopes at NASA's Deep Space Network facility near Canberra (Australia), the Hartebeesthoek telescope (South Africa; currently inoperative), TIGO (Chile), and O'Higgins (Antarctica). A typical \mbox{($u$-$v$)-coverage} for a source at \mbox{$-43^\circ$ declination} is shown in Fig.1, providing an angular resolution of about \mbox{$0.9$ -- $0.6\,\mathrm{mas}$} at 8.4\,GHz. At each frequency one $24$-hour epoch including 24 sources is observed approximately every two months.\\\\
The initial TANAMI sample of 43 sources has been defined combining a radio selected flux-density limited subsample and a $\gamma$-ray selected subsample of known and candidate $\gamma$-ray sources based on results of \textit{CGRO/EGRET}. It contains all known radio sources from the catalogue of Stickel et al. (1994) above a limiting radio flux density of $S_\mathrm{5\,GHz} > 2\,\mathrm{Jy}$ which have a flat spectrum between $2.7\,$GHz and $5\,$GHz. Since 2008 November, new \textit{Fermi} LAT-detected AGN have added to the monitoring program so that the sample now consists of all known radio- and $\gamma$-ray bright AGN of the Southern Hemisphere.\\\\
For a better understanding of jet structure and dynamics, multiwavelength observations are fundamental. The emission of blazars extends over the whole electromagnetic spectrum and a close connection between $\gamma$-ray and radio emission is suggested. The analysis of radio-jet properties of LAT-detected TANAMI sources helps to answer key questions concerning beaming factors of superluminal motion, origin of high energy production, and the connection between $\gamma$-ray flares and jet-component ejections.\\\\
First results of the TANAMI program are discussed by Ojha et al. (submitted to A\&A) and summarized below.\\ 
The middle and right panels of Fig. 1 show examples of $8.4$ and $22\,$GHz images of the $\gamma$-ray bright BL Lac object PKS B\,0521$-$365, respectively. After 3 months of observations, $\sim$28\% of TANAMI sources have been detected by \textit{Fermi}-LAT. The redshift distribution of quasars peaks at $z \sim 1.5$, while BL Lacs and galaxies are distributed at $z < 0.4$ resembling the distributions for the bright $\gamma$-ray AGN seen by LAT (Abdo et al. 2009). The five most distant and most luminous TANAMI sources and the 9 most luminous radio loud AGN have not been detected by LAT in the first three months of operation. There is no significant difference in the brightness temperature distributions of LAT detected and non-detected sources. Quasars have the higher $T_\mathrm{B}$-values; 13 of the initial 43 sources have a maximum $T_\mathrm{B}$ below the equipartition value, 29 below the inverse Compton limit, putting about a third of the sources above this limit. Furthermore our analysis suggests that $\gamma$-ray bright AGN have larger opening angles than those not detected by LAT, when excluding galaxies.\\
\begin{figure}[h]
\begin{minipage}{0.33\linewidth}
  \includegraphics[width=0.91\linewidth]{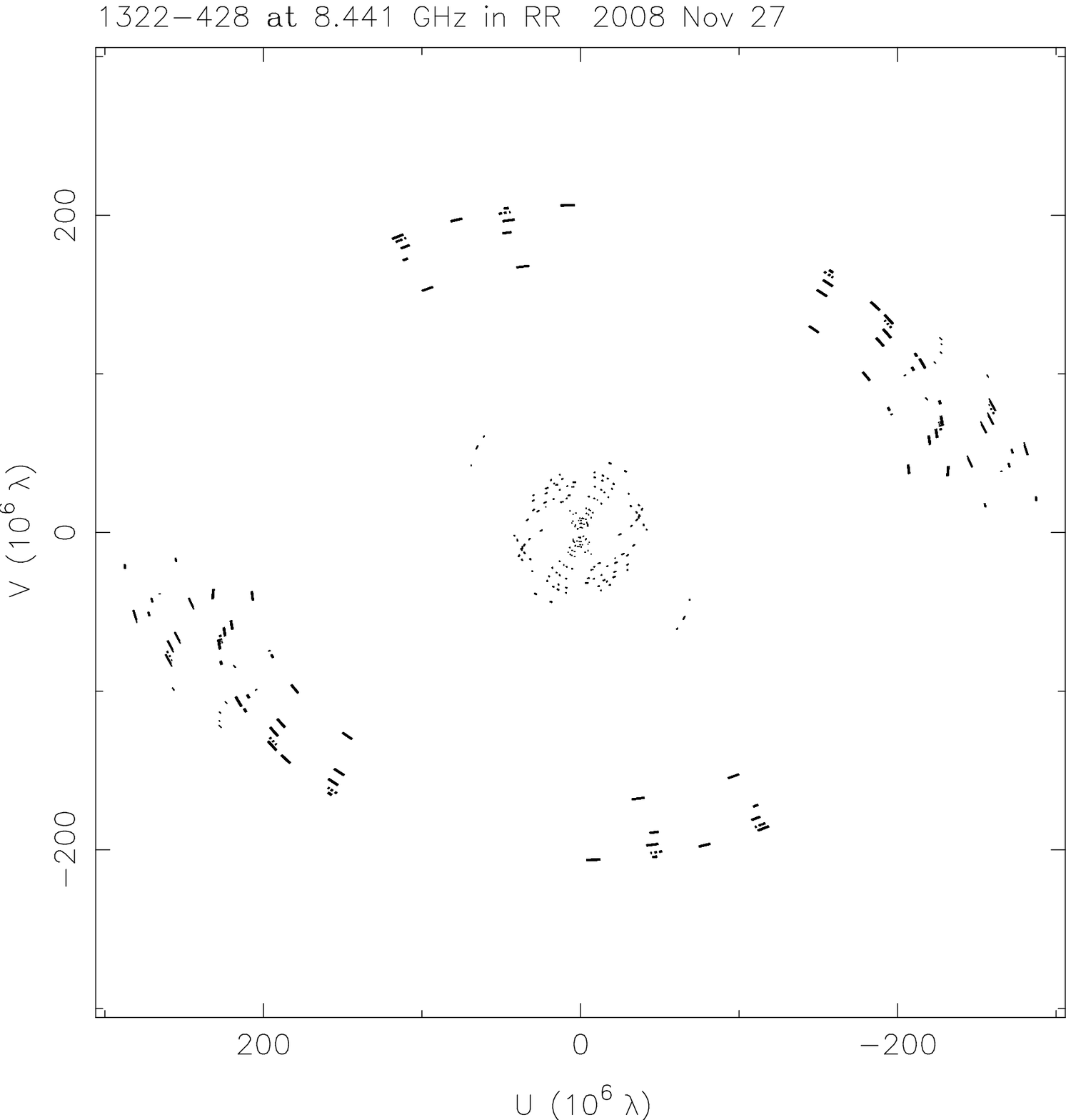}
\end{minipage}\hfill
\begin{minipage}{0.33\linewidth}
  \includegraphics[width=0.89\linewidth]{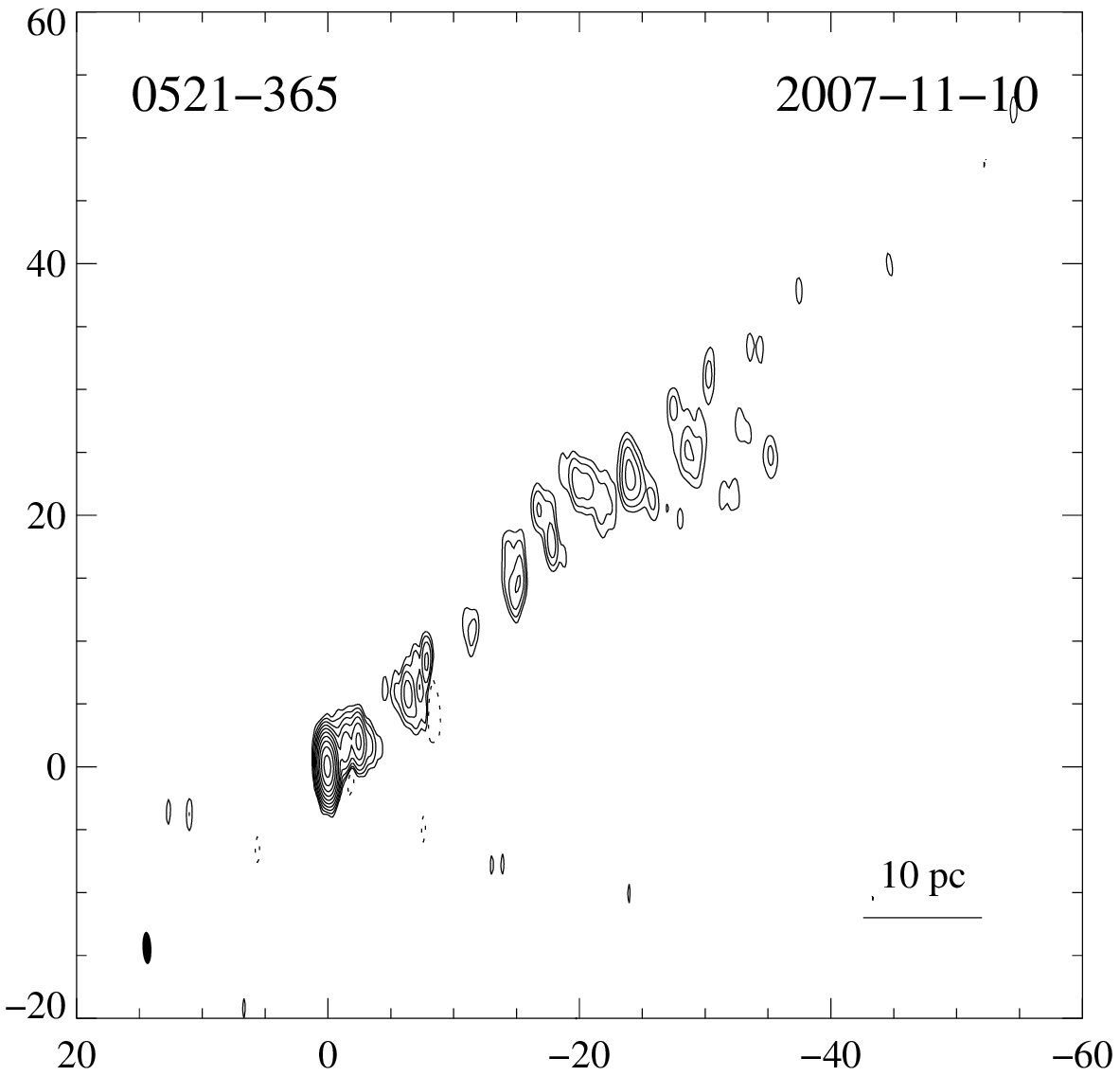}
\end{minipage}\hfill
\begin{minipage}{0.33\linewidth}
  \includegraphics[width=0.89\linewidth]{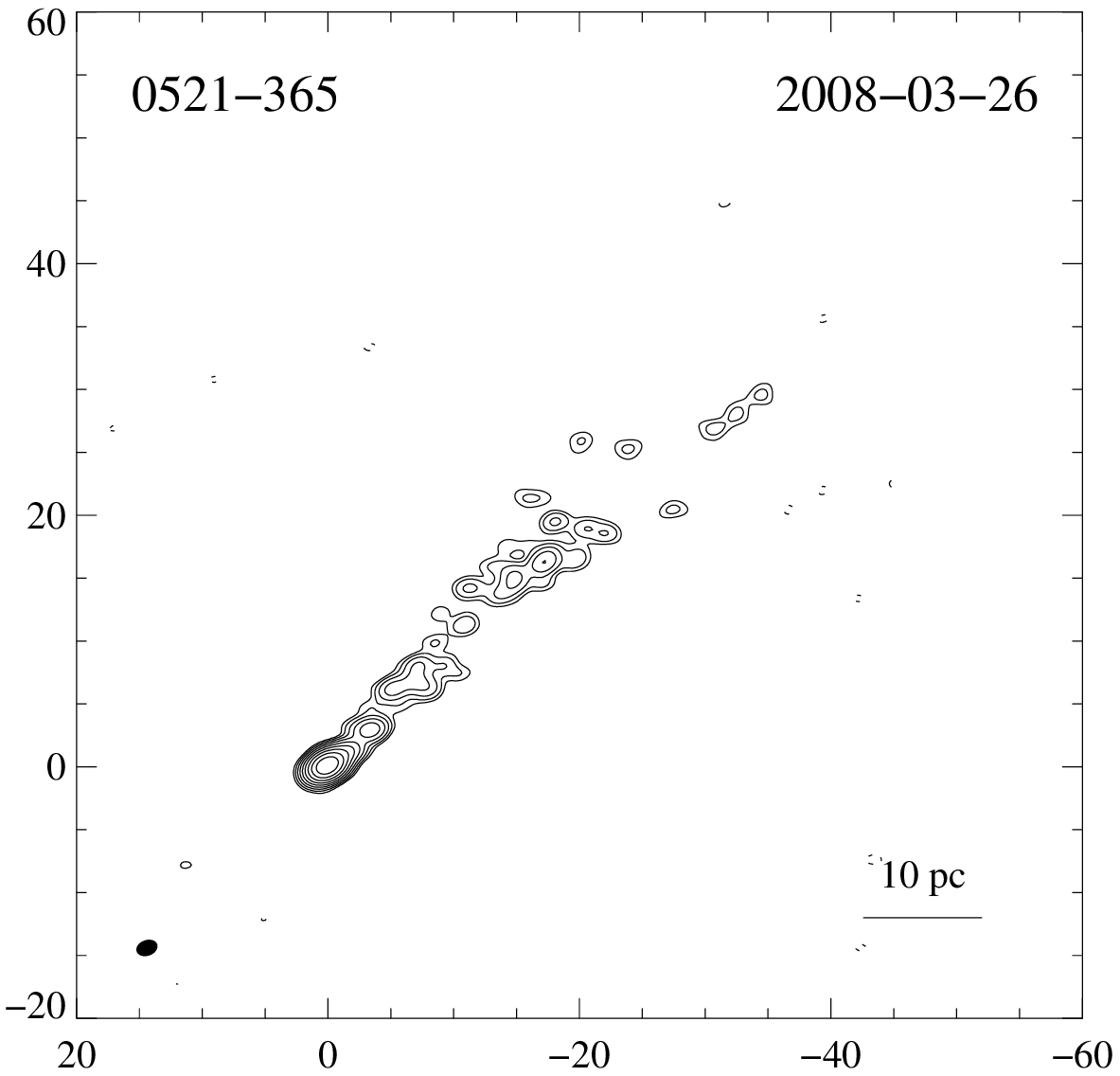}
\caption{Typical ($u$-$v$)-coverage at 8.4\,GHz for Centaurus A (left); 8.4 GHz (middle) and 22 GHz (right) images of PKS B\,0521$-$365. The units of the images are in milliarcseconds and measured relative to the core at each frequency; the lowest contours are at 1.5\,mJy and 1.8\,mJy, respectively.}
\end{minipage}
\end{figure}

\newpage
\textbf{Full author list}:\\\\
C. M\"uller$^{1}$, M. Kadler$^{1,2,3}$, R. Ojha$^{4,5}$, M. B\"ock$^{1}$,
R. Booth$^{6}$, M. S. Dutka$^{7}$, P. Edwards$^{8}$, A. L. Fey$^{4}$, L. Fuhrmann$^{9}$,
H. Hase$^{10}$, S. Horiuchi$^{11}$, D. L. Jauncey$^{8}$, K. J. Johnston$^{4}$, U. Katz$^{1}$,
M. Lister$^{12}$, J. E. J. Lovell$^{13}$, C. Pl\"otz$^{14}$, J. F. H. Quick$^{6}$, E. Ros$^{9,15}$,
G. B. Taylor$^{16}$, D. J. Thompson$^{17}$, S. J. Tingay$^{18}$, G. Tosti$^{19, 20}$,
A. K. Tzioumisk$^{8}$, J. Wilms$^{1}$ and J. A. Zensus$^{9}$ 
\quote{$^1$ Dr. Remeis-Sternwarte \& ECAP, Sternwartstrasse 7, 96049 Bamberg, Germany\\
$^2$ CRESST/NASA Goddard Space Flight Center, Greenbelt, MD 20771, USA\\
$^3$ USRA, 10211 Wincopin Circle, Suite 500 Columbia, MD 21044, USA\\
$^4$ USNO, 3450 Massachusetts Ave., NW, Washington DC 20392, USA\\
$^5$ NVI, Inc., 7257D Hanover Parkway, Greenbelt, MD 20770, USA\\
$^6$ Hartebeesthoek Radio Astronomy Observatory, PO Box 443, Krugersdorp 1740, South Africa\\
$^7$ The Catholic University of America, 620 Michigan Ave., N.E.,  Washington, DC 20064\\
$^8$ ATNF, CSIRO, PO Box 76, Epping, NSW 1710, Australia\\
$^9$ MPIfR, Auf dem H\"ugel 69, 53121 Bonn, Germany\\
$^{10}$ BKG, Univ. de Concepcion, Casilla 4036, Correo 3, Chile\\
$^{11}$ CDSC, PO Box 1035, Tuggeranong, ACT 2901,  Australia\\
$^{12}$ Dept. of Physics, Purdue Univ., 525 Northwestern Avenue, West Lafayette, IN 47907, USA\\
$^{13}$ School of Math. \& Phys., Private Bag 37, Univ. of Tasmania, Hobart TAS 7001, Australia\\
$^{14}$ BKG, Geodetic Observatory Wettzell, Sackenrieder Str. 25, 93444 Bad Koetzting, Germany\\
$^{15}$ Dept. d'Astronomia i Astrof\'{\i}sica, Univ. de Val\`encia, 46100 Burjassot, Val\`encia, Spain\\
$^{16}$ Dept. of Physics and Astronomy, Univ. of New Mexico, Albuquerque NM, 87131, USA\\
$^{17}$ ASD, NASA Goddard Space Flight Center, Greenbelt, MD 20771, USA\\
$^{18}$ Curtin Institute of Radio Astronomy, Curtin Univ. of Tech., Bentley, WA, 6102, Australia\\
$^{19}$ INFN, Sezione di Perugia, 06123 Perugia, Italy\\
$^{20}$ Dipartimento di Fisica, Universit\`a degli Studi di Perugia, 06123 Perugia, Italy\\
}
\end{document}